\documentclass[twocolumn]{IEEEtran}
\usepackage{epsf}

\newcommand\etal{{\em et al. }}

\def\BibTeX{{\rm B\kern-.05em{\sc i\kern-.025em b}\kern-.08em
    T\kern-.1667em\lower.7ex\hbox{E}\kern-.125emX}}

\begin{document}

\title{A Generalization of the Maximum Noise Fraction Transform}

\author{Christopher Gordon\thanks{C. Gordon is with the School of Computer Science and
Mathematics at the University of Portsmouth in the UK. E-mail:christopher.gordon@port.ac.uk}
}
\newpage

\maketitle

\abstract{A generalization of the maximum noise fraction 
(MNF) transform is proposed. Powers of each band are included as new bands before the MNF
transform is performed. 
 The generalized MNF (GMNF) is shown to perform better than 
the MNF on a time dependent airborne electromagnetic (AEM) data
filtering problem.
\footnote{\bf 
Copyright (c) 2000 Institute of Electrical and Electronics 
Engineers.   Reprinted from [IEEE Transactions on Geoscience and
Remote Sensing, Jan 01, 2000, v38, n1 p2, 608].
This material is posted here with permission of the IEEE.  Internal or 
personal use of this material is permitted.  However, permission to 
reprint/republish this material for advertising or promotional purposes or 
for creating new collective works for resale or redistribution must be 
obtained from the IEEE by sending a blank email message to 
pubs-permissions@ieee.org.
By choosing to view this document, you agree to all provisions of the 
copyright laws protecting it.}
}


\section{Introduction}

\PARstart{T}{he} maximum noise fraction (MNF) transform was introduced by Green
\etal 
[1].  
It is similar to the principle component transform 
[2]
in that
it consists of a linear transform of the original data. However, the 
MNF transform orders the bands in terms of noise fraction. 

One application of the MNF transform  is noise filtering of
multivariate data 
[1].
The data is MNF transformed,
the high noise fraction bands are filtered and then the reverse transform
is performed.

We show an example where the MNF noise removal adds artificial features
due to the nonlinear relationship between the different variables of the data.
A polynomial generalization of the MNF is introduced which removes this problem.

In Section 
II
 we summarize the MNF procedure. The problem
data set is introduced in Section
III
and the MNF is 
applied to it. In Section
IV,
the generalized MNF transform
is explained and applied. The conclusions are given in Section 
V.

\section{The Maximum Noise Fraction (MNF) Transform}
\label{sec:MNF_outline}
In this section we  define the MNF transform and list some of its
properties. For further details the reader is referred to 
Green \etal  
[1]
and Switzer and Green 
[3].
A good review is also given by Nielsen
[4].
A reformulation of the MNF transform as the noise-adjusted principle component
(NAPC) transform was given by Lee \etal    
 [5].
An efficient method of computing the MNF transform is given by Roger 
[6].

Let
\[
     Z_i(x), \quad i = 1,\ldots, p
\]
be a multivariate data set with $p$ bands and with $x$ giving the position of
the sample. The means of $Z_i(x)$ are assumed to be zero. The data can
always be made to approximately satisfy this assumption by subtracting the 
sample means.
An additive noise model is assumed:
\[
        Z(x) = S(x) + N(x)
\]
where $Z^{T}(x) = \{Z_1(x), \ldots, Z_p(x)\}$ is the corrupted signal and
$S(x)$ and $N(x)$ are the uncorrelated signal and noise components of $Z(x)$.
The covariance matrices are related by:
\[
        \mbox{Cov}\{Z(x)\} = \Sigma = \Sigma_S + \Sigma_N
\]
where $\Sigma_N$ and $\Sigma_S$ are the noise and signal covariance matrices.

The noise fraction of the $i$th band is defined as
\[
        \mbox{Var} \{N_i(x)\} / \mbox{Var} \{Z_i(x)\}.
\]
The maximum noise fraction transform (MNF) results in a new $p$ band 
uncorrelated data
set which is a linear transform of the original data:
\[
        Y(x) = A^{T}Z(x).
\]
The linear transform coefficients, $A$, are found by solving the eigenvalue
equation:
\begin{equation}
        A \Sigma_N \Sigma^{-1}  = \Lambda A
\label{eq:MNF_coef}
\end{equation}
where $\Lambda$ is a diagonal matrix of the eigenvalues, $\lambda_i$.
The noise fraction in $Y_i(x)$ is given by $\lambda_{i}$. By convention
the $\lambda_i$ are ordered so that $\lambda_1 \ge \lambda_2 \ge \ldots
\ge \lambda_p$. Thus the MNF transformed data will be arranged in bands of
{\em decreasing\/} noise fraction. 
The proportion of the noise variance described by the first $r$ MNF 
bands is given by
\[
        \frac{\sum_{i = 1}^r \lambda_i}{\sum_{i = 1}^p \lambda_i}.
\]
The eigenvectors are normed so that $A^{T}\Sigma A$ is equal to an identity
matrix. 

The advantages of the MNF transform over the PC transform are that it is 
invariant to linear transforms on the data and the MNF transformed bands are 
ordered by noise fraction.

The high noise fraction bands can be filtered and then the transform reversed.
This can lead to an improvement in the filtering results because the high noise
fraction bands should contain less signal that might be distorted by the filtering.
Examples of this approach have been given by Green \etal  
[1],
Nielsen and Larsen
[7]
and Lee \etal  
[5].

An extreme version of MNF filtering is based on excluding the effects of the first
$r$ components. 
That is $r$ is chosen so as to include only bands with high
enough noise ratios.
 This can be achieved by:
\begin{equation}
\label{eq:mnf_filter}
        Z^{*}(x) = (A ^ {-1})^{T} R A^{T} Z(x)
\end{equation}
where $Z^{*}(x)$ is the filtered data and $R$ is an identity matrix with the
first $r$ diagonal elements set to zero.
Thus eliminating the effect of one or more of the MNF bands produces a
filtered data set which is a linear transform  of the original data. 
This MNF based filter uses interband correlation to remove noise.

In order to use Equation 
(1)
to compute $A$, $\Sigma_N$ has
to be known. Nielsen and Larsen 
[7]
have given four different ways of estimating $N(x)$.
 They all rely on the data being spatially correlated. A simple method
 for computing $N(x)$ is by
\begin{equation}
\label{eq:noise_estimation}
     N(x) = Z(x) - Z(x + \delta)
\end{equation}
where $\delta$ is an appropriately determined step length. We are effectively
assuming
\[
    S(x) = S(x + \delta).
\]
To the extent that this is not true, the estimate of $N(x)$ is in error.

When this method of noise estimation is used, the MNF transform is equivalent
to the min / max autocorrelation factor transform 
[3].

\section{Airborne Electromagnetic Data}
\label{sec:AEM_data}
We test the MNF filtering methodology on a flight line produced by
SPECTREM's time dependent airborne electromagnetic (AEM) system. Background information
on this AEM system has been explained by Leggatt 
[8].
A multiband image can be formed by consecutive flight lines but usually
each flight line is examined separately.

Fig. 1 shows a  flight line of data, consisting of the 
 seven windowed AEM X band spectra. All seven bands are displayed stacked
above each other.
\begin{figure}[p]
\leavevmode
\centering
\epsffile[168 131 422 456]{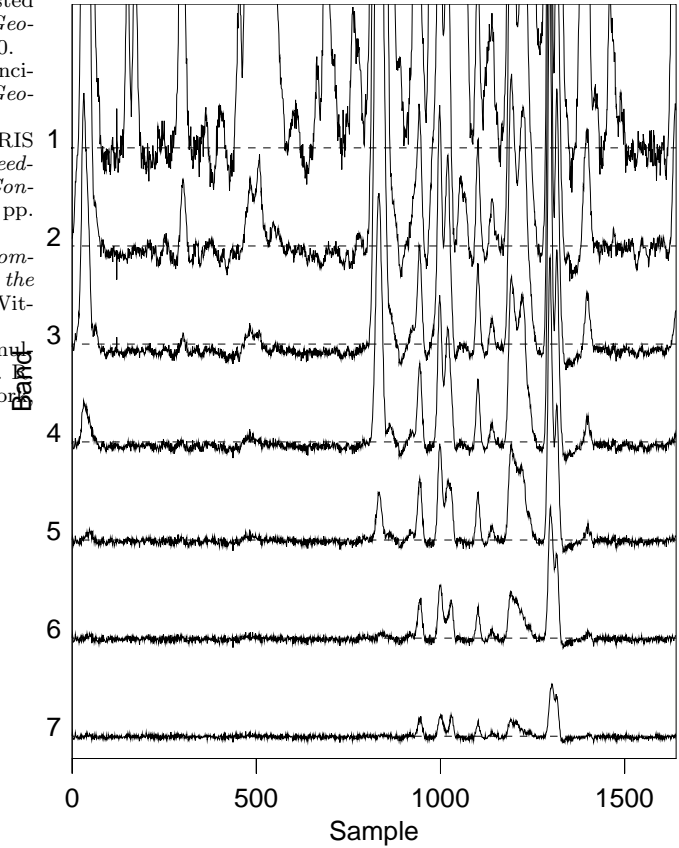}
\caption{Unfiltered AEM data. Bands 1 to 7 are shown.
 The band number of each spectrum is  labelled to
the left of the spectrum. 
 The dotted line of each spectrum marks the 
 zero amplitude for that spectrum.}
\end{figure}
The amplitude of a band at a particular point is proportional to the 
 vertical distance of the 
spectrum  from its corresponding zero amplitude reference (dotted) line.
 Neighbouring points along a line are
responses from neighbouring points on the ground. The higher band numbers
are associated with greater underground depths.

Ore bodies are often associated with small features in the higher bands. 
Analysis can be made easier by filtering the spectra.
Because this data set has substantial interband correlation, the MNF filtering
methodology can be used.

Fig. 2 (b)
 shows the MNF filtering of the spectra in Fig. 1. Only the 
last three bands (i.e. 5, 6 and 7)  and a portion of the flight line are shown.
\begin{figure}[p]
\epsfysize=14cm
\leavevmode
\centering
\epsffile[167 137 408 713]{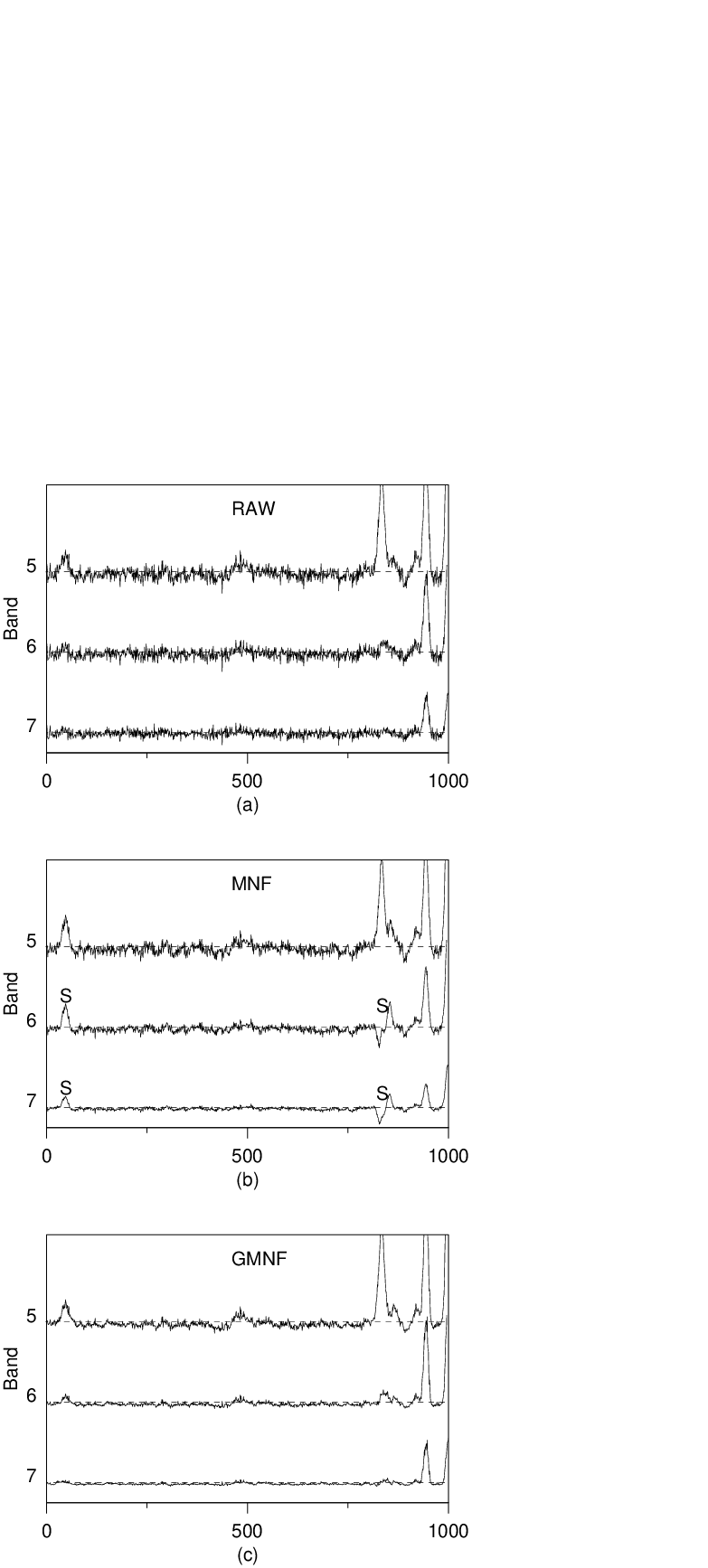}
\caption{
  A comparison of the MNF and GMNF filtering methods. Only a portion
        of the flight line for bands 5, 6 and 7 is shown for each figure.
        The sample number is displayed on the horizontal axis of each subplot.
        (a) Unfiltered AEM data. (b) MNF filtered AEM data. The `S' symbols mark
        parts of the data where spurious features have been introduced by the MNF filtering. (c) GMNF filtered
        AEM data.
}
\end{figure}
The noise was estimated by taking the difference in neighboring pixels, as
in Equation 
(3).
 The data were filtered by excluding  the first two
MNF bands which accounted for approximately 86\% of the noise fraction.
Although the noise has been reduced, spurious features have been added, indicated by `S'.
Excluding
only the last MNF component does not significantly reduce the magnitude of the 
spurious features and does almost no noise reduction. 

As seen in Equation 
(2),
 the MNF filtered data is  composed
from a linear function of the original data. Fig. 3
shows a plot of $Z_1(x)$ against $R_{Z_1}(x)$, where $R_{Z_1}(x)$ is 
the difference between $Z_1(x)$ and a least squares regression of
$Z_1(x)$ based on all the other bands.
\begin{figure}[p]
\leavevmode
\centering
\epsffile[131 136 408 407]{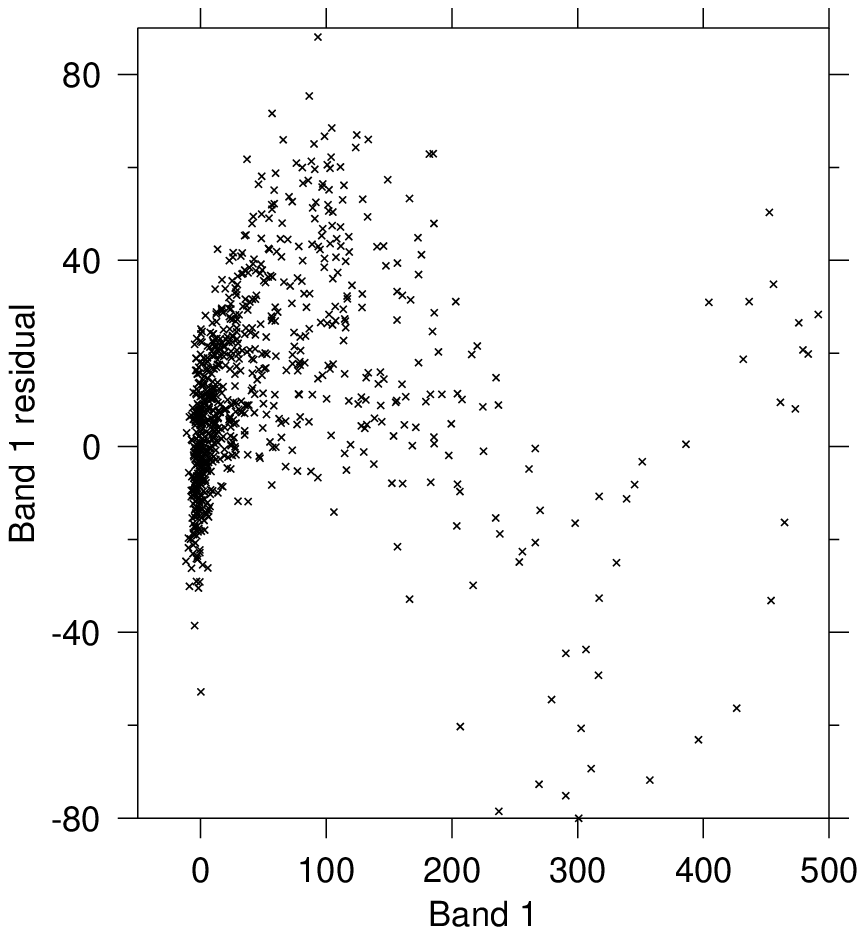}
\caption{
A plot of the residual of a linear regression of band 1 based
on bands 2 to 7, versus band 1 values.
}
\end{figure}
The clear pattern of the residuals plotted in Fig. 3
is evidence that the relationship between $Z_1(x)$ and the other bands is not
linear. Similar patterned residuals were found for residual
plots based on the other bands. In the next section we show how the linear
assumption can be relaxed.

\section{The Generalized Maximum Noise Fraction Transform}
\label{generalized_MNF}
From the discussion in the previous section it  appears that 
using a linear filter is too restrictive
for this data set. Gnanadesikan 
[9]
 proposed a
 generalization of the principle component transform.
 Powers of the original bands were appended to the
data set as new bands. For example, $p$ new bands can be created by appending
the square of each band to the original data set. Thus each generalized
principle component would be a polynomial, as opposed to linear, function of
all the bands in the original data set.

The same procedure can be applied to generalize the MNF transform.
More formally,
a new data set, $Z'(x)$, can be created by appending up to $q$ powers
of the original data set:
\begin{eqnarray*}
        Z'(x) & = & \{Z_1(x), Z_2(x), \ldots, Z_p(x), Z_1 ^2(x), Z_2^2(x), 
                \ldots, Z_p^2(x),\ldots, \\
              &  & \mbox{} Z_1(x)^q,\ldots, Z_p^q(x)\}.
\end{eqnarray*}
We are assuming that the $Z_i(x)$ have zero means.
Cross terms, such as $Z_1(x) Z_2(x)$ can also be appended. The rest of
the methodology remains unchanged.

From Equation 
(2),
each band of the generalized MNF (GMNF)
filtered data can be seen to be,
\[
        Z^{*}_i(x) =  \sum_{j = 1}^p \sum_{k = 1} ^ q F_{i,j + (k -1)p}
                     Z_j^k(x)
\]
where $F_{i,j + (k - 1)p}$ is the element in row $i$ and column $j + (k - 1)p$
 of the filter matrix:
\[
        F =(A ^ {-1})^{T} R A^{T}
\]
Thus the GMNF transform leads to a polynomial filter.

To apply the GMNF filter to the data in Fig. 1, the GMNF
transform was applied with  powers of up to order 6 for each band
appended to the original data. Cross terms were found to make little difference
to the result and so were not included. The first 15 of the 42 GMNF components,
contributing approximately 80\% of the noise fraction, were eliminated.

Fig. 2(c) shows the GMNF filtered AEM data. A comparison with  the MNF filtered
data (Fig. 2(b)) shows that for GMNF filtered data , 
the noise reduction is greater and spurious features
are much less evident.

\section{Conclusion}
\label{sec:conclusion}
We have proposed a generalized maximum noise fraction transform (GMNF) that
is a polynomial as opposed to linear transform.
The GMNF was applied  to  filtering a test AEM data set. It was
found to remove more noise while adding less artificial features than the
MNF based filter.

Implementing the GMNF is a simple extension of the MNF implementation.
 Software written for the MNF transform can be be used for the GMNF 
transform without any modification.

\bibliographystyle{IEEE}


\end{document}